\documentclass[preprint,aps]{revtex4}

\usepackage{graphicx}
\usepackage{dcolumn}
\usepackage{bm}
\usepackage{amsfonts}
\usepackage{amssymb}

\begin{document}
\title{Rayleigh-B\'{e}nard convection with phase changes in a Galerkin model}
\author{Thomas Weidauer}
\affiliation{Institut f\"ur Thermo- und Fluiddynamik, Technische Universit\"at Ilmenau,
             Postfach 100565, D-98684 Ilmenau, Germany}
\author{Olivier Pauluis}
\affiliation{Courant Institute of Mathematical Sciences, New York University, 251 Mercer Street, New York, NY 10012-1185, USA}
\author{J\"org Schumacher}
\affiliation{Institut f\"ur Thermo- und Fluiddynamik, Technische Universit\"at Ilmenau,
             Postfach 100565, D-98684 Ilmenau, Germany}
\date{\today}             
\begin{abstract}
The transition to turbulence in Rayleigh-B\'{e}nard convection with phase changes 
and the resulting convective patterns are studied in a three-dimensional Galerkin model. 
Our study is focused to the conditionally unstable regime of moist convection in which 
the stratification is stable for unsaturated air parcels and unstable for saturated parcels. We perform a
comprehensive statistical analysis of the transition to convection that samples the dependence 
of attractors (or fixed points) in the phase space of the model on the dimensionless parameters. Conditionally unstable convection can be initiated either from a fully unsaturated 
linearly stable equilibrium or a fully saturated linearly unstable equilibrium. 
Highly localized moist convection can be found in steady state, in oscillating
recharge-discharge regime or turbulent in dependence on the aspect ratio and the degree of 
stable stratification of the unsaturated air. Our phase space analysis predicts parameter ranges
for which self-sustained convective regimes in the case of 
subcritical conditional instability can be observed. The observed regime transitions for moist convection bear some similarities to transitions to turbulence in simple shear flows.
\end{abstract}
\maketitle
\section{Introduction}
Rayleigh-B\'{e}nard (RB) convection, in which a fluid is heated from below and cooled from above, is one of the most comprehensively studied flow configurations, either in the regime of transition to convection \cite{Busse1978} and pattern formation \cite{Bodenschatz2000} or in the fully turbulent case \cite{Ahlers2009,Lohse2010}. In nearly all these studies the working fluid is in a single phase. Moist convection combines Rayleigh-B\'{e}nard convection and changes between two fluid phases, such as the gaseous and liquid ones. It is characterized by the internal release of latent heat for condensation and an increase of the efficient heat conductivity. This was demonstrated in a recent laboratory convection experiment with ethane \cite{Zhong2009} and in numerical simulations \cite{Oresta2009}. It is ubiquitous throughout the atmosphere of the Earth through the formation of clouds \cite{Emanuel1994,Stevens2005}, but has also its importance in many industrial applications reaching from miniaturized heat exchangers \cite{Dhir1998} to huge cooling towers in power plants \cite{Hensley2006}.

Condensation and evaporation are not only passive processes. The interaction of phase change and (turbulent) fluid flow results in a very complex dynamics intrinsic to the formation of moist plumes and clouds. The thermodynamics of phase change is highly nonlinear, but can be considerably simplified through the use of a piecewise linear approximation of the equation of state.
This approach was suggested by Kuo \cite{Kuo1961} and Bretherton \cite{Bretherton1987,Bretherton1988} and has been further developed recently \cite{Pauluis2010}. Such simplified thermodynamics enables a systematic way to investigate the dependence of convection on additional internal heating and to bridge the results directly to classical RB convection. This was done recently in direct numerical simulations (DNS) of turbulent convection with phase changes 
\cite{Weidauer2010,Schumacher2010,Pauluis2011}. However, even for a simplified two-phase convection model the space of dimensionless system parameters is at least five-dimensional and consequently prevents a detailed exploration by DNS.

In this work, we present a systematic study of the dependence of moist 
Rayleigh-B\'{e}nard convection on the dimensionless parameters. Our investigation is focused 
to the transition to moist convection. In order to perform such a study, we significantly reduce the 
degrees of freedom in comparison to a fully resolved DNS. For this purpose, we construct a 
three-dimensional Galerkin model based on Fourier modes. This mode reduction limits our analysis to Rayleigh numbers of ${\cal O}(10^4)$ but makes it possible to monitor parameter dependencies systematically and to sample the structure of the underlying phase space by a statistical analysis 
in the spirit of recent investigations of the transition to turbulence in wall-bounded shear flows 
\cite{Eckhardt2007,Eckhardt2007a}.

Our focus is on moist convection in the so-called {\em conditionally unstable (CU) regime}. Dry air parcels are then stable with respect to vertical lifts while saturated air parcels are unstable and can rise in convective plumes.  We address the following questions: How is the transition to convection in the CU regime characterized? How does it depend on the degree of saturation of the initial equilibrium state of the system? Which phase space structure is associated  with the transition to moist convection?

Models with reduced degrees of freedom are a powerful tool to study weakly nonlinear regimes in fluids. Several approaches to develop reduced models have been proposed, such as Galerkin models based on proper orthogonal decomposition \cite{Smith2005}, reduced waveset approximations of turbulent flows \cite{Eggers1991,Grossmann1992} or models in which only a least set of the largest-scale \cite{Saltzman1962,Lorenz1963} or a few dynamically relevant modes \cite{Waleffe1997,Eckhardt1999,Moehlis2004} are captured. In dry Rayleigh-B\'{e}nard convection such models provided a deeper understanding of the transition to convection \cite{Saltzman1962,Lorenz1963} and the weakly nonlinear dynamics right above the onset of convective motion \cite{Malkus1958,Schlueter1965}. Ogura and Phillips \cite{Ogura1962} and later Shirer and Dutton \cite{Shirer1979} applied systematic mode expansion to the moist convection case and set up minimal models.

The transition to convective turbulence in dry Rayleigh-B\'{e}nard convection proceeds via a sequence of bifurcations of increasing spatial and temporal complexity. Convection sets in with stationary roll patterns right above the critical threshold, bifurcates to other stationary states or to smooth time-dependent flow patterns for higher Rayleigh numbers and eventually becomes turbulent (see e.g. Ref. \cite{Busse1978} for a comprehensive review). Dry convection is thus one of the standard applications of normal mode analysis to study linear stability and to seek for stationary nonlinear solutions at the transition threshold. 
Such normal mode analysis is only partly applicable for special configurations of moist convection since derivatives of the buoyancy field with respect to thermodynamic state variables have a discontinuity at the phase boundary, albeit the buoyancy field (which drives the fluid motion) itself is continuous.

Bretherton analyzed the onset of moist convection by demonstrating the existence of linear two-dimensional growing modes \cite{Bretherton1987} and by studying their behavior in the weakly nonlinear regime of convection \cite{Bretherton1988}. These studies showed that 
moist convection preferably develops in isolated saturated rising plumes separated from each other by 
broad unsaturated subsiding regions. This particular property of conditionally unstable convection was already predicted by Bjerknes \cite{Bjerknes1938} in a strikingly simple qualitative model. It is well-known from field measurements such as in Refs. \cite{Malkus1953,Malkus1964} and comprehensive large-eddy simulations of atmospheric moist convection containing additional physical processes such as radiative transfer, ice formation  and precipitation \cite{Siebesma1998,Bretherton2005}.

The purpose of our study is to extend Brethertons analysis of the conditionally unstable regime in various directions. It is based on the set of RB equations which we developed in \cite{Pauluis2010} and which are similar to \cite{Bretherton1987}. First, we consider the three-dimensional case which has not been discussed in \cite{Bretherton1987,Bretherton1988}. Second, the studies in 
\cite{Bretherton1987,Bretherton1988} are limited to a situation where the initial quiescent equilibrium is always exactly at the saturation line. This means that the layer is saturated but no condensed water is present in the initial profile, albeit subsequent convective motions can lead to further condensation. We will refer this regime to as the Kuo-Bretherton (KB) case. We extend conditionally unstable convection in both the {\em subcritical} case, corresponding to an initially unsaturated layer in which saturated air parcels may still be convective unstable, and the {\em supercritical} case, corresponding to an initially fully saturated layer in which condensed water is present at all levels \cite{Pauluis2011}. 
This is achieved by increasing or decreasing the degree of stratification of the dry air compared to the parameters in the 
Kuo-Bretherton case. Third, we will perform a statistical analysis of the transition to turbulence that 
unravels the underlying phase space structure of our dynamical system in the sub- and supercritical cases. In particular, the subcritical case is known to be stable for small perturbations. A key issue here will be to determine for which ranges of the dimensionless parameters the self-sustained convective regimes can be obtained or not. We show that this transition -- similar to the transition to turbulence
in fundamental shear flows, such as a pipe or plane Couette flow -- depends on the shape and the
amplitude of the perturbation, a result of the competition between two co-existing attracting sets in 
phase space which are separated by a complex hyperplane.

The outline of this paper is as follows. The model equations and the numerical implementation of the Fourier-Galerkin approximation are explained in Section 2. Section 3 presents the two equilibrium states followed by section 4 that summarizes our analysis of the transition behavior and the evolving convection states. Finally, we will give a short summary and outlook to future work.
\section{Galerkin model}
\subsection{Moist Rayleigh-B\'{e}nard convection}
We now review briefly our model for moist Rayleigh-B\'{e}nard convection. A detailed derivation and discussion  can be found in Refs. \cite{Pauluis2010,Weidauer2010,Schumacher2010}. In the case of thermal convection, fluid motions are driven by a buoyancy force $B$ that is added to the momentum balance equation to account for the variations of density in the fluid. The buoyancy field $B$ is given by the following relation \cite{Pauluis2008}
\begin{equation}
B(S,q_v,q_l,p)=-g\, \frac{\rho(S,q_v,q_l,p)-\overline{\rho}}{\overline{\rho}} \,, \label{buo0}
\end{equation}
with $g$ being the gravity acceleration, $\overline{\rho}$ a mean density, $p$ the pressure, $S$ the entropy and $q_v$, $q_l$ the specific humidity of water vapor and liquid water. The following simplifications are made in the model. The first one is that the air parcels are in local thermodynamic equilibrium, which means that the two specific humidities can be combined to a specific humidity for the total water, a new state variable $q_T$ which remains constant for adiabatic non-precipitating processes:
\begin{equation}
q_T=q_v+q_l\,.
\end{equation}
This reduces the number of prognostic variables to three. The second simplification is to apply the Boussinesq approximation to the system which in effect replaces the pressure dependence of $B$ by a height dependence, i.e. $B = B(S,q_T,z)$.

The dependence of the buoyancy field on the two remaining variables of state, $S$ and $q_T$, and the vertical coordinate $z$ still contains the full thermodynamics of phase changes. One can approximate $B(S,q_T,z)$ as a piecewise linear function of $S$ and $q_T$ around the phase boundary between unsaturated gaseous and saturated liquid phases. This step preserves the discontinuity of partial derivatives of $B(S,q_T,z)$ with respect to $q_T$ and $S$ at the phase boundary (and thus the release of latent heat). The last two steps limit the applicability of the model to shallow convection where the thickness of the atmospheric layer remains moderate.

Since $B$ is a linear function of the two variables of state $S$ and $q_T$, we can introduce two new prognostic fields, a {\em dry buoyancy field} $D$ and a {\em moist buoyancy field} $M$ as linear combinations of $S$ and $q_T$. An air parcel is saturated and liquid water is present  when \cite{Pauluis2010}
\begin{equation}
M-D+N_s^2 z\ge 0\,.
\end{equation}
$N_s$ is the Brunt-Vaisala frequency which is defined as $N_s^2=g(\Gamma_d-\Gamma_s)/T_{ref}$. Here, $\Gamma_d$ and $\Gamma_s$ are the dry and moist adiabatic lapse rates and $T_{ref}$ is a reference temperature, e.g. the temperature at the bottom plane. All these assumptions lead to a 
formula of the buoyancy field $B(M,D,z)$ which is given by
\begin{equation}
B({\bf x},t)=\max\left( M({\bf x},t), D({\bf x},t)-N^2_s z \right) \,. \label{auftrieb}
\end{equation}
The saturation condition (\ref{auftrieb}) is a nonlinear relation which can be easily carried out in numerical simulations. On the basis of (\ref{auftrieb}), we can define the amount of
condensate by
\begin{equation}
q_l({\bf x},t)=M({\bf x},t)- D({\bf x},t)+N^2_s z \ge 0 \,. \label{wassergehalt}
\end{equation}
Consequently, all points $q_l\ge 0$ belong to moist plumes with droplets or clouds. 

The dry and moist buoyancy fields can be decomposed into
\begin{eqnarray}
D({\bf x},t)&=& \overline{D}(z)+D^{\prime}({\bf x},t)\nonumber \label{dequi}\\
                  &=& D_0+\frac{D_H-D_0}{H}z+D^{\prime}({\bf x},t)\\
M({\bf x},t)&=& \overline{M}(z)+M^{\prime}({\bf x},t)\nonumber \label{mequi}\\
                  &=& M_0+\frac{M_H-M_0}{H}z+M^{\prime}({\bf x},t)\,.
\end{eqnarray}
Here, $D_0$ and $M_0$ are prescribed values at $z=0$ respectively, $D_H$ and $M_H$ at $z=H$. The variations about the linear equilibrium profiles of both fields, $D^{\prime}$ and $M^{\prime}$, have to vanish at $z=0$ and $H$. Equation (\ref{auftrieb}) can now be transformed into
\begin{equation}
B=\overline{M}(z) + \max\left( M', D' + \overline{D}(z) -\overline{M}(z)  -N^2_s z \right) \,. \label{auftrieb2}
\end{equation}
To obtain a dimensionless version of the model equations, one has to define characteristic scales. These are the height of the layer $H$, the velocity $U_f=\sqrt{H (M_0-M_H)}$, the time $T=H/U_f$, the characteristic kinematic pressure $U_f^2$, and the buoyancy difference $M_0-M_H$. The dimensionless equations contain three nondimensional parameters, the Prandtl number $Pr=\nu/\kappa$ and the dry and moist Rayleigh numbers $Ra_D$ and $Ra_M$ which are given by
\begin{equation}
Ra_D =\frac{H^3 (D_0 - D_H )}{\nu \kappa},\;\;\;Ra_M =\frac{H^3 (M_0 - M_H )}{\nu \kappa}\,.
\end{equation}
Here $\nu$ is the kinematic viscosity and $\kappa$ is the diffusivity of the buoyancy fields. In addition,
 two more parameters arise from the saturation condition (\ref{auftrieb2}): 
 the  Surface Saturation Deficit ($SSD$) and the Condensation in Saturated Ascent ($CSA$)
\begin{equation}
SSD = \frac{D_0 - M_0}{M_0 - M_H}, \;\;\; CSA = \frac{N^2_s H}{M_0 - M_H}\,.
\end{equation}
The parameter $SSD$ determines the degree of saturation of the  air parcels at the bottom 
plane $z=0$. For the rest of the work, we set $D_0=M_0$ and thus $SSD=0$. The second new parameter, $CSA$, is proportional to the amount of water formed in a saturated parcel adiabatically rising from the bottom to the top. The buoyancy $B$ in dimensionless form is given by
\begin{equation}
B = \max \left( M^{\prime}, D^{\prime} +  SSD + \left(1-\frac{Ra_D}{Ra_M}-CSA\right) z \right) \label{GG05}\,.
\end{equation}
 We can decompose the buoyancy field in the last equation in a mean contribution $\overline{B}(z)$ which can be added to the kinematic pressure $p$ and a fluctuating field $B^{\prime}$, i.e. $\partial_z 
p +B =\partial_z \tilde{p}+B^{\prime}$ . The dimensionless equations are then
\begin{eqnarray}
&\partial_t {\bf u}+({\bf u} \cdot\nabla){\bf u} = - \nabla \tilde{p} + \sqrt{\frac{Pr}{Ra_M}} \nabla^2{ \bf u} + B^{\prime} {\bf e}_z \label{GG01} \\
&\nabla \cdot \bf {u} =0 \label{GG02} \\
&\partial_t D^{\prime}+ ({\bf u}\cdot\nabla) D^{\prime} =  \frac{1}{\sqrt{PrRa_M}} \nabla^2 D^{\prime} +\frac{Ra_D}{Ra_M} u_z \label{GG03} \\
&\partial_t M^{\prime}+ ({\bf u}\cdot\nabla) M^{\prime} =  \frac{1}{\sqrt{PrRa_M}} \nabla^2 M^{\prime} + u_z \, . \label{GG04}
\end{eqnarray}
Here ${\bf u}$ is the velocity field. Together with (\ref{GG05})  this forms a closed system 
of equations that describes moist Rayleigh-B\'{e}nard convection. Beside $SSD=0$, we 
set $Pr=0.7$ and $CSA=4/3$ in this work.

An alternative choice for the two non-dimensional parameters which are related to the 
phase change is given by
\begin{eqnarray}
CW_0 &=&-\frac{SSD}{CSA} \nonumber  \,\\
CW_H &=& 1+CW_0 +\frac{Ra_D / Ra_M -1}{CSA} \nonumber \, .
\end{eqnarray}
A positive value of either $CW_0$ or $CW_H$ corresponds to cloud water (CW) present at the bottom or top. A negative value stands for a water deficit. Here $CW_0=0$, while $CW_H$ will vary with a change of $Ra_D$.
\subsection{Galerkin approximation}
The model equations are solved in a rectangular box of height 1 and width $AR$, which is the aspect ratio of the domain. In the horizontal directions $x$ and $y$ we use periodic boundary conditions. In the vertical direction $z$ we apply free-slip boundary conditions at $z=0,1$
\begin{equation}
u_z = D^{\prime} = M^{\prime} =0\,,\;\;\;\;\; \frac{\partial u_x}{\partial z} = \frac{\partial u_y}{\partial z} =0\,. \label{BC}
\end{equation}
Galerkin approximations are frequently used to investigate dynamics that mainly includes a few modes \cite{Eckhardt1999,Moehlis2004,Tong2002,Shirer1979}. We will use this method here for comprehensive statistical and parametric studies that would not be possible with direct numerical simulations in which all modes to the smallest flow scale are included. Our Galerkin approximation is based on a truncated Fourier series. For a simple implementation of the boundary conditions we 
double the box from $z=-1$ to $z=1$. For a wavevector ${\bf n}=(n_x,n_y,n_z) \in \mathbb{Z}^3$ we introduce the following notation
\begin{equation}
\hat{{\bf n}} = \left( \frac{2 \pi i n_x}{AR}, \frac{2 \pi i n_y}{AR}, \pi i n_z \right) \,.
\end{equation}
Note that the aspect ratio is included in this definition and does not emerge directly in the following equations. Every field is expanded in a Fourier series. For example the velocity field is given 
by
\begin{equation}
{\bf u} \left( {\bf x},t \right) = \sum_{ {\bf n} \in \mathbb{Z}^3} {\bf u} \left( {\bf n},t \right ) e^{ \hat{{\bf n}} {\bf x}} \,.
\end{equation}
The following set of ordinary differential equations results from (\ref{GG01}) -- (\ref{GG04})
\begin{eqnarray}
\partial_t {\bf u}({\bf n},t) &=& -\sum_{{\bf p}+{\bf q}={\bf n}} ({\bf u} ({\bf p},t)\cdot \hat{{\bf q}}) {\bf u} ( {\bf q},t) + B^{\prime}({\bf n},t){\bf e}_z \nonumber \\
&& -\hat{{\bf n}} \tilde{p}({\bf n},t) +\sqrt{\frac{Pr}{Ra_M}} \hat{{\bf n}}^2 {\bf u}({\bf n},t)\label{GGF01} \\
{\bf u} ({\bf n},t) \cdot \hat{{\bf n}} &=& 0 \label{GGF02} \\
\partial_t D^{\prime}({\bf n},t) &=& \frac{\hat{{\bf n}}^2 }{\sqrt{PrRa_M}}  D^{\prime}({\bf n},t) +\frac{Ra_D}{Ra_M} u_z({\bf n},t) \nonumber \\
&&-\sum_{{\bf p}+{\bf q}={\bf n}} ({\bf u} ({\bf p},t)\cdot \hat{{\bf q}}) D^{\prime} ( {\bf q},t) \label{GGF03} \\
\partial_t M^{\prime}({\bf n},t) &=& \frac{\hat{{\bf n}}^2}{\sqrt{PrRa_M}} M^{\prime}({\bf n},t) +u_z({\bf n},t) \nonumber \\
&& -\sum_{{\bf p}+{\bf q}={\bf n}} ({\bf u} ({\bf p},t)\cdot \hat{{\bf q}}) M^{\prime} ( {\bf q},t) \label{GGF04} \,.
\end{eqnarray}
Since the number of modes is small compared to a direct numerical simulation, the convolution sums in Eqns. (\ref{GGF01})-(\ref{GGF04}) can be calculated directly. Taking the divergence of (\ref{GGF01}) and using (\ref{GGF02}) we get the following equation for the pressure,
\begin{eqnarray}
-\hat{\bf n} \tilde{p}({\bf n},t) &=&\frac{1}{\hat{\bf n}^2} \hat{\bf n} \sum_{{\bf p}+{\bf q}={\bf n}} (({\bf u}({\bf p},t)\cdot \hat{\bf q})({\bf u}({\bf q},t) \cdot \hat{\bf n}) \nonumber \\
&& - \hat{\bf n} i \pi n_z B^{\prime}({\bf n},t) ) \,.
\end{eqnarray}
Since all quantities have to be real in physical space it follows that 
\begin{eqnarray}
{\bf u} (-{\bf n},t)&=&{\bf u}^{\ast} ({\bf n},t) \\
M^{\prime} (-{\bf n},t)&=& M^{\prime\ast} ({\bf n},t) \\
D^{\prime} (-{\bf n},t)&=& D^{\prime\ast} ({\bf n},t) \,,
\end{eqnarray}
where the asterisk stands for complex conjugate. The free-slip boundary conditions 
(\ref{BC}) are implemented by an additional symmetry with respect to the plane $z=0$ 
for $u_x$, $u_y$ and antisymmetry for $u_z$, $M^{\prime}$ and $D^{\prime}$. This results to
\begin{eqnarray}
u_x (x,y,-z,t) &=& u_x(x,y,z,t) \\
u_z (x,y,-z,t) &=& -u_z(x,y,z,t) \, .
\end{eqnarray}
In Fourier space this requires
\begin{eqnarray}
u_x (n_x,n_y,-n_z,t) &=& u^{\ast}_x(n_x,n_y,n_z,t) \\
u_z (n_x,n_y,-n_z,t) &=& -u^{\ast}_z(n_x,n_y,n_z,t) \,.
\end{eqnarray}
Bretherton \cite{Bretherton1987} noticed already that a further simplification can be made by 
using that $M^{\prime}$ and $D^{\prime}$ obey the same linear equations (except for a different 
driving amplitude) and boundary conditions. We can consequently use
\begin{equation}
D^{\prime}=\frac{Ra_D}{Ra_M}M^{\prime} \, .
\end{equation}
Thus only one buoyancy field has to be simulated. The Fourier coefficients $B^{\prime}({\bf n},t)$ are calculated in the following way. First $B^{\prime}$ is evaluated in physical space on a grid. Then the coefficients
\begin{equation}
B^{\prime} ({\bf n},t)= \frac{1}{V}\int_{V}B^{\prime}({\bf x},t) e^{\hat{n} {\bf x}} d{\bf x}
\end{equation}
are computed using 4th-order rule for the evaluation of integrals \cite{Hermann2}. The time stepping is done with a third order Runge-Kutta scheme by von Heun \cite{Hermann1}. The choice of the
wave vectors is a compromise between computational costs and incorporating the large-scale flow aspects. Note also that the saturation condition (\ref{GG05}) will generate a fair amount of scale interactions. 
Furthermore remember that $B^{\prime}$ is not differentiable at the phase 
boundary, but  approximated by continuous differentiable functions. Therefore a certain number of modes is necessary to get an appropriate approximation. All Galerkin model runs at $AR=3,4$ are 
done with the wave vectors $(|n_x| \le 5,|n_y|\le 5,|n_z|\le 5)$ if not stated otherwise. Mode (0,0,0)
is always excluded. For larger aspect ratios more wavevectors are taken. These are in total
975 independent degrees of freedom since one velocity component can be determined by the incompressibility condition.
\begin{figure}
\begin{center}
\includegraphics[angle=0,scale=0.6,draft=false]{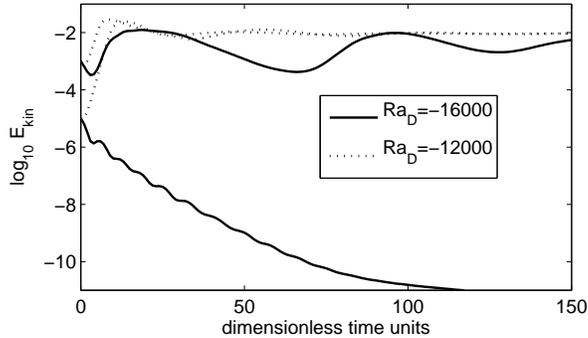}
\caption{Examples for runs in the subcritical (solid lines) and the supercritical cases (dotted lines). 
In each regime we perturb the equilibrium either with a finite initial perturbation or an infinitesimal perturbation. The transition to convection or the return to the equilibrium are displayed by the corresponding kinetic energy versus time.}
\label{Bsp1}
\end{center}
\end{figure}
\section{Conditionally unstable equilibria}
In the absence of fluid motion, the profiles of $D$ and $M$ will be linear across the layer as given in Eqns. (\ref{dequi}) and (\ref{mequi}). The choice of the four amplitudes, $D_0$, $D_H$, $M_0$, and $M_H$, causes different classes of equilibrium states.
Conditionally unstable layers are defined as layers in which clear air is stably stratified, i.e. $D_H>D_0$ and thus $Ra_D<0$. Moist air is unstably stratified, i.e. $M_H<M_0$ and thus $Ra_M>0$. To classify the CU equilibria of a moist convection layer
one defines the {\it Convectively Available Potential Energy} ($CAPE$) \cite{Lorenz1955} by
\begin{equation}
CAPE=\int_0^1 [B(\overline{D}(0) , \overline{M}(0) ,z )-B(\overline{D}(z) , \overline{M}(z) ,z )]dz\,.
\end{equation}
The first term describes the potential energy of the air parcels that start at the bottom. The second term stands for the potential energy of the background equilibrium profile. A necessary condition for the onset of convection is that $CAPE>0$. This is equivalent to ($SSD=0$)
\begin{equation}
CAPE >0 \Leftrightarrow CSA  > -\frac{Ra_D}{Ra_M}. 
\end{equation}
In the following, we will discuss two cases of conditionally unstable convection and investigate their stability properties.

The quiescent equilibrium configuration of the convection layer is fully subsaturated if the linear profiles satisfy
\begin{equation}
\overline{M}(z)<\overline{D}(z)-N_s^2 z \nonumber \,,
\end{equation} 
for all $z\in [0,1]$. This results together with (\ref{GG05}) to
\begin{equation}
CSA  > -\frac{Ra_D}{Ra_M} > CSA -1  \,.
\end{equation}
We will refer convection which arises from this equilibrium case to as the {\em subcritical CU regime}. 
In this regime, the linear profile is unsaturated, and small perturbations cannot change the saturation of air parcels. In effect, small perturbations experience the stable stratification of the dry buoyancy field $D$, and the diffusive equilibrium solution is linearly stable. Finite perturbations however can trigger moist convection out of the equilibrium (see next section). This is demonstrated in Fig. \ref{Bsp1} which compares the evolution of the kinetic energy for two perturbations of different initial amplitudes. Note that if the stable stratification of the dry buoyancy field is too strong, $CAPE$ cannot be positive. This is the absolute stability threshold, a necessary (but not sufficient) condition to initiate moist convection. It defines one interval boundary of the subcritical regime. 

The other interval boundary of the subcritical regime is set by the so-called Kuo-Bretherton (KB) case \cite{Kuo1961,Bretherton1987}. This case is established if the degree of stable stratification of $D$ is 
reduced until $\overline{M}(z)=\overline{D}(z)-N_s^2 z$ corresponding to
 \begin{equation}
 -Ra_D/Ra_M = CSA - 1 \,. 
 \end{equation}
 The whole domain is held in  an equilibrium at the saturation threshold. This case has been extensively studied by Bretherton \cite{Bretherton1987,Bretherton1988} who demonstrated the existence of exponentially growing solutions in the form of isolated saturated plumes and investigated their behavior in the weakly nonlinear regime.

Once the stratification is further reduced, small-amplitude perturbations to the equilibrium can trigger moist convection as demonstrated by the dotted curves in Fig. \ref{Bsp1}. The quiescent equilibrium is fully saturated for
\begin{equation}
\overline{M}(z)>\overline{D}(z)-N_s^2 z\,. \nonumber
\end{equation} 
We will refer convection triggered from this equilibrium state to as the {\em supercritical CU regime} because the equilibrium is linearly unstable, which corresponds to non-dimensional parameters satisfying the condition
\begin{equation}
 CSA - 1 > -\frac{Ra_D}{Ra_M}  > 0 \,.
\end{equation}
Note that when the value of $CSA$ is less than $1$, there is no supercritical CU regime 
due to the fact that saturation of the entire layer would require a positive value of the dry 
Rayleigh number $Ra_D$ for these values of the moist Rayleigh number $Ra_M$ and 
$CSA$.  The case of both buoyancy fields unstably stratified, corresponding to $Ra_D > 0$ 
and $Ra_M > 0$, is also possible and has been discussed in detail in \cite{Pauluis2010,Schumacher2010,Weidauer2010}. In the following section we will discuss 
both CU regimes.
\section{Transition to moist convection}
\subsection{Subcritical regime}
\subsubsection{Statistical analysis of the transition to convection}
\begin{figure}
\begin{center}
\includegraphics[angle=0,scale=0.6,draft=false]{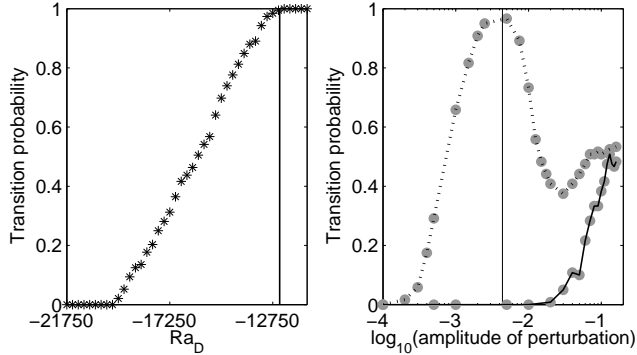}
\caption{Statistical analysis of the transition to moist convection. We show the transition probabilities to obtain a convective solution B. Left: Dependence on $Ra_D$ for fixed magnitude of initial perturbation, $E_{kin}$ and $E_{ M^{\prime}}$. The vertical line marks the dry Rayleigh number corresponding with the KB case. Right: Amplitude dependence of the transition probability. Solid line is for transition from convective state B to diffusive state A, dotted line from state A to state B. The vertical line marks the kinetic energy of state B for $Ra_D=-1.60\times 10^4$. All data are for $AR=4$ and $Ra_M=3.73\times 10^4$ which is for the subcritical CU regime.} 
\label{Statistical}
\end{center}
\end{figure}

In the subcritical regime, the linear equilibrium solution is stable. After an initial perturbation and burst of convection,
the layer can either return to the diffusive equilibrium state (denoted as A in the following) or ends up in a convective state with fluid motion (denoted as B) for the same set of dimensionless parameters $Ra_M$, $Ra_D$ and $AR$ when a finite
perturbation of different shape or amplitude is applied. Such a behavior is similar to the transition to turbulence in wall-bounded shear flows \cite{Eckhardt2007}. To investigate this behavior systematically, we fixed the aspect ratio to 4 and performed a statistical investigation in the spirit of those in shear
flows \cite{Schmiegel2000,Faisst2004}. Therefore, we took 192 randomly selected perturbations among the wave vectors with wave numbers $|n_x|=1,2$, $|n_y|=1,2$, and $|n_z|=1,2$ for the velocity field and the moist buoyancy field. The initial kinetic energy $E_{kin}=\langle {\bf u}^2\rangle_V/2$ and the variance of the moist buoyancy field $E_{M^{\prime}}=\langle M^{\prime\,2}\rangle_V/2$ were the same for all 192 runs. The result is shown in Fig. \ref{Statistical} (left). For $Ra_D$ chosen in the supercritical regime, state  B is always obtained. The probability  to end in steady moist convection is unity. For $Ra_D$ taking values in the subcritical regime,  the probability of a relaxation to state B decreases to zero with decreasing $Ra_D$. Note that the value of $Ra_D$ with zero probability to obtain B is still significantly larger then the dry Rayleigh number which is associated with the threshold of $CAPE=0$. This would 
result to $Ra_D=-4.98 \times 10^4$ for the present example.

We also determined the smallest value of $Ra_D$ to observe a (stationary) convection state B. State B for $Ra_D$ close to zero transition probability is taken as an initial condition for a new simulation with a slightly reduced $Ra_D$. This case is advanced in time as long as it needs to relax anew. The procedure can be continued iteratively and the relaxation to the new (stationary) state increases in time for decreasing $Ra_D$. In this way we found state B appearing at Rayleigh numbers as small as $Ra_D=-1.99\times 10^4$.

\subsubsection{Dependence on initial perturbation amplitude}
Results on the amplitude dependence of the transition at a fixed set of system parameters are shown in Fig. \ref{Statistical} (right). We started therefore either from state A or B and determined the probability to reach state B or A, respectively, by varying the amplitude of the initial perturbation. The moist buoyancy field $M^{\prime}$ is not perturbed. One can observe that the convective state B is very stable with respect to  small perturbations (solid line) and becomes unstable when the kinetic energy of the perturbation exceeds a threshold which is marked as a solid vertical line in the figure. In contrast, the diffusive state A (dotted line) behaves differently and shows a finite probability to switch to the convective state B over nearly the whole range of
perturbations.
\begin{figure}
\begin{center}
\includegraphics[angle=0,scale=0.6,draft=false]{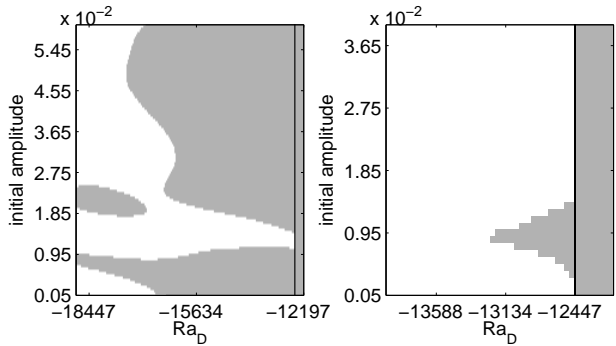}
\caption{Transition to moist convection as a function of the perturbation amplitude and the dry Rayleigh number. Left: $AR=4$. Here, 120 different values for $Ra_D$ and 192 different amplitudes are taken. Right: $AR=3$. Here, 40 different values of $Ra_D$ and 94 different perturbation amplitudes are taken. Gray stands for reaching state B out of the equilibrium, white for a return to state A. The vertical line marks the KB case at the border between sub- and supercritical regimes. Each simulation was run for 300 dimensionless time units which is more than a diffusive time $t_d=H^2/\nu$.} 
\label{Transition}
\end{center}
\end{figure}

The observation suggests a more systematic monitoring of the plane which is spanned
by the perturbation amplitude and $Ra_D$, as done for example in a Galerkin model for a plane
shear flow \cite{Moehlis2004}. Now the shape of the perturbation is fixed. The amplitude of the perturbation and the dry Rayleigh number are varied to cover both CU regimes. The detailed scan of the parameter plane is shown in Fig. \ref{Transition} for aspect ratios of 3 (right) and 4 (left). Perturbations that cause a return to A are in white and those that initiate a convection state are in gray.

The figure illustrates nicely that the boundary between the two attracting regions is complex if one samples the phase space, even for the moderate Rayleigh numbers discussed here. This is for example obvious from the isolated island that can be seen in the left panel of Fig. \ref{Transition} for $AR=4$. The observation is similar to what has been observed in low-dimensional shear flow models. An additional property of shear flows is the finite lifetime of transient states \cite{Faisst2004,Schmiegel2000} which we did not observe for the present system. Once moist convection is initiated it remains sustained, except at the boundary of the basin of attraction of state B (see Fig. \ref{Edge} and subsequent text).

\subsubsection{Structure of convection states in the subcritical regime}
\begin{figure}
\begin{center}
\includegraphics[angle=0,scale=0.6,draft=false]{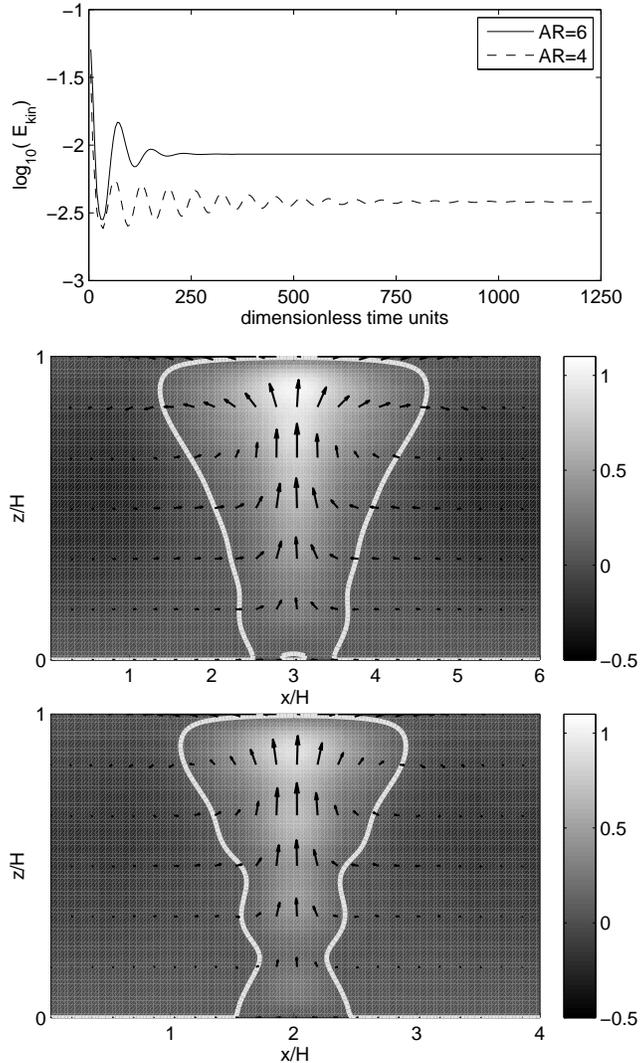}
\caption{Examples for stationary single moist plume (or cloud) at $Ra_D=-1.50 \times 10^4$ and $Ra_M=3.73\times 10^4$. The upper plot shows the kinetic energy vs. time that becomes eventually stationary. The solid line corresponds to $AR$=6 and the mid plot, the dashed one to $AR$=4 an the lower plot. The mid and bottom panels show centered cross section contour plots in the $x-z$ plane of the final (stationary) states. We show the velocity field vectors in the plane and the contours of liquid water content $q_l$ as defined in (\ref{wassergehalt}). The bright solid line in the mid and bottom figures marks the boundary with $q_l$=0. All points inside this boundary with $q_l>0$ belong to the cloud. Points with $q_l<0$ are outside the cloud and stand for a water deficit.} 
\label{Bsp2}
\end{center}
\end{figure}
When convection is initiated, state B  takes the shape of a single localized steady moist plume (or cloud) at small aspect ratio. It is characterized by a strong saturated upward motion inside the cloud balanced by a weak downward motion outside -- a well known feature of moist convection in this 
regime as already mentioned in the introduction and e.g. in \cite{Emanuel1994}. The position of the single cloud in the box differs from case to case, but the shape and the flow structure are always the same. It is observed for moderate magnitudes of $Ra_D$ and $Ra_M$ at different aspect ratios as seen in  Fig. \ref{Bsp2}. Similar solutions have been reported by Bretherton \cite{Bretherton1987,Bretherton1988} in his two-dimensional simulations of the nonlinear evolution.

\begin{figure}
\begin{center}
\includegraphics[angle=0,scale=0.6,draft=false]{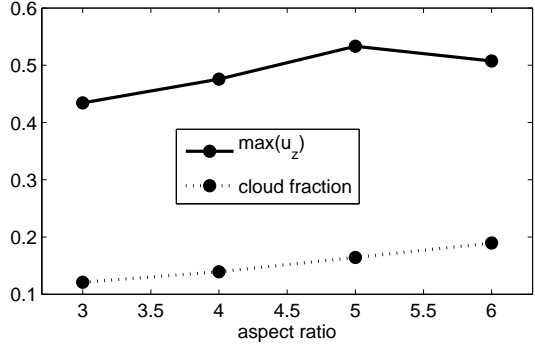}
\caption{Maximum vertical velocity inside the moist plume and fraction of saturated air with $q_l \geq 0$ as a function of the aspect ratio. Here, $Ra_D=-1.50\times 10^4$ and $Ra_M=3.73\times 10^4$.} 
\label{aspect}
\end{center}
\end{figure}

For $AR=2$, the convective state B was not detected at all. The size of the box is then too small to provide enough space for the dry subsidence outside the cloud and thus to form a stable cloud pattern. For the remaining cases, i.e. $AR=3,4,5,6$, a single cloud appears in the box. Neighboring clouds are stabilizing each other (recall the periodic boundary conditions in $x$ and $y$). States B for several $AR$ do not differ qualitatively in their overall structure as supported by the mid and bottom panels of Fig. \ref{Bsp2}. With increasing $AR$, the clouds become more and more smooth and grow in diameter relatively to the aspect ratio. Quantitatively, the velocity magnitude and the cloud fraction (all points with $q_l>0$) of the layer increase with the aspect ratio, as displayed in Fig. \ref{aspect}.

For even larger aspect ratios, such as $AR=7$ and $8$, we still observe a single cloud in the domain, but now temporal fluctuations of all fields can evolve. The stabilizing mechanism from neighboring clouds is too weak. It is mentioned here that the number of wave vectors and the computational grid have been always adjusted with growing $AR$ in the Galerkin model. Multiple clouds \textnormal{with periodic distance apart} have also been observed in \cite{Bretherton1988}.

\subsubsection{Equilibrium and moist plume as coexisting attractors}
One possible explanation for the observed transition behavior in the subcritical regime is that two coexisting attractors (or fixed points) which are associated with equilibrium state A and (stationary) convective solution B. Both attractors and their associated basins of attraction in phase space will vary with changes in the parameters. While the basin of B will grow with increasing $Ra_D$, the basin of state A will shrink. The coexistence of two attractors is also supported by the observation that initiated
convection states do not decay after a finite time.

To confirm that there is a small attracting set in the phase space close to both, states A and B, we proceed as follows for one particular parameter set. Both fixed point solutions, denoted as ${\bf X}^{\ast}_A$ and ${\bf X}^{\ast}_B$, are slightly perturbed. The time advancement is started with initial conditions taken from a sphere in phase space around both attracting states, i.e., ${\bf X}^j_{A,B}(t=0)={\bf X}^{\ast}_{A,B}+\delta{\bf X}^j$ with $j=1,...,1016$. They are randomly selected on this sphere with $|\delta{\bf X}^j|=5\times 10^{-4}$. All cases relaxed either to states A or B. The dry Rayleigh number was fixed at $Ra_D=-1.85\times 10^4$ for this investigation, and the aspect ratio is $AR = 4$. Our study is not a rigorous proof, but provides evidence for our present picture of phase space. Repeating this procedure in the supercritical regime at $Ra_D=-10^4$, we find that the system escapes from state A for all 1016 initial conditions, while state B remains attractive as before. In  other words, the saturated equilibrium state (state A) is now linearly unstable. 

\subsubsection{Tracking the edge state between both attractors}
\begin{figure}
\begin{center}
\includegraphics[angle=0,scale=0.6,draft=false]{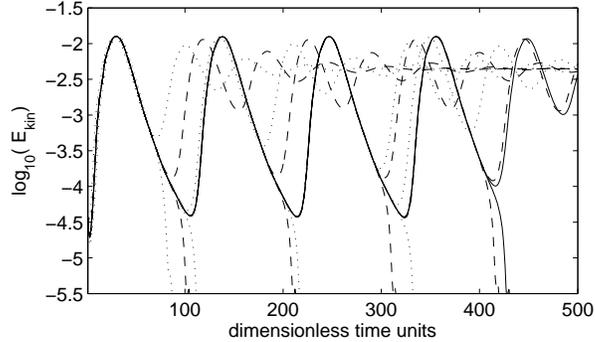}
\caption{Edge state tracking for $AR=4$ and $Ra_D=-1.60\times 10^4$. We plot the kinetic energy versus time. Shown are always the two runs at the given refinement level that enclose the edge state. To distinguish the refinement levels the corresponding kinetic energy curves are plotted dotted and dashed in alternating sequence. The last iteration that was possible with double precession corresponds to the solid curves.}
\label{Edge}
\end{center}
\end{figure}

\begin{figure}
\begin{center}
\includegraphics[angle=0,scale=0.6,draft=false]{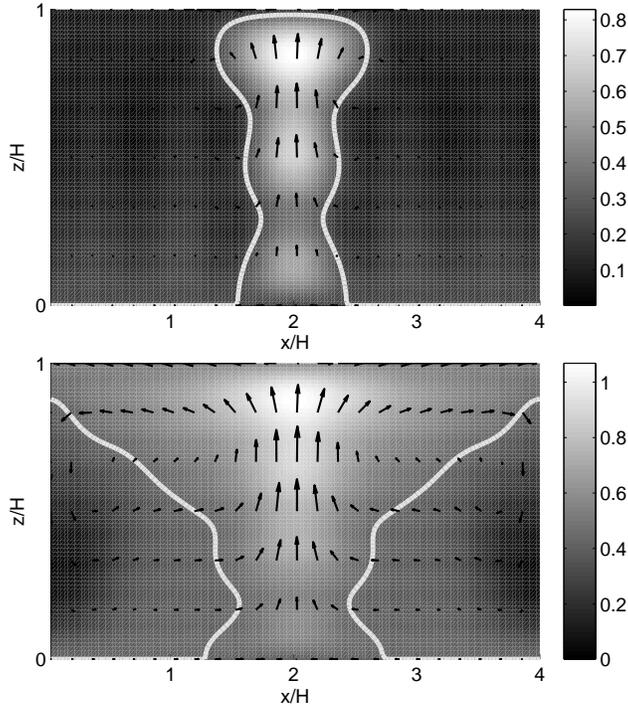}
\caption{Velocity and liquid water content in the $x-z$ plane for $AR$=4. The bright line marks again $q_l=0$. Top figure: $Ra_D=-1.96\times 10^4$ in the subcritical regime. Bottom figure: $Ra_D=-7.85\times 10^3$ in the supercritical regime. For both cases $Ra_M=3.73\times 10^4$.} 
\label{Bsp4}
\end{center}
\end{figure}

In the following, we explore the boundary between both basins of attraction in more detail. Solutions which can be found on this boundary are denoted as edge states \cite{Skufca2006,Schneider2007}. We used an edge tracking method which is similar to Ref. \cite{Schneider2008}. The boundary that separates both basins of attraction must be somewhere between states A and B, and can be obtained as a linear combination of the two fixed points:
\begin{equation}
{\bf X}(t=0)=\lambda  {\bf X}^{\ast}_A +(1-\lambda) {\bf X}^{\ast}_B\,,
\end{equation}
with $\lambda \in [0,1]$. Because state A is the origin it reduces simply to $\lambda {\bf X}^{\ast}_B$. We divided the interval [0,1] into 65 equidistant points and detected the bin with the value of $\lambda$ 
that separates state A from state B. This procedure is refined successively and the refined values of $\lambda$ are used to start a new run at $t=0$. Results
for $Ra_D=-1.60\times 10^4$ are shown in Fig. \ref{Edge}. The detected edge state is a periodic orbit. It is a cloud fixed at one position that first grows to a cloud with shape and flow structure very similar to state B (see Figs. \ref{Bsp2} and \ref{Bsp4} (top)). Afterwards it collapses to a small cloud pen, but never vanishes.

\begin{figure}
\begin{center}
\includegraphics[angle=0,scale=0.25,draft=false]{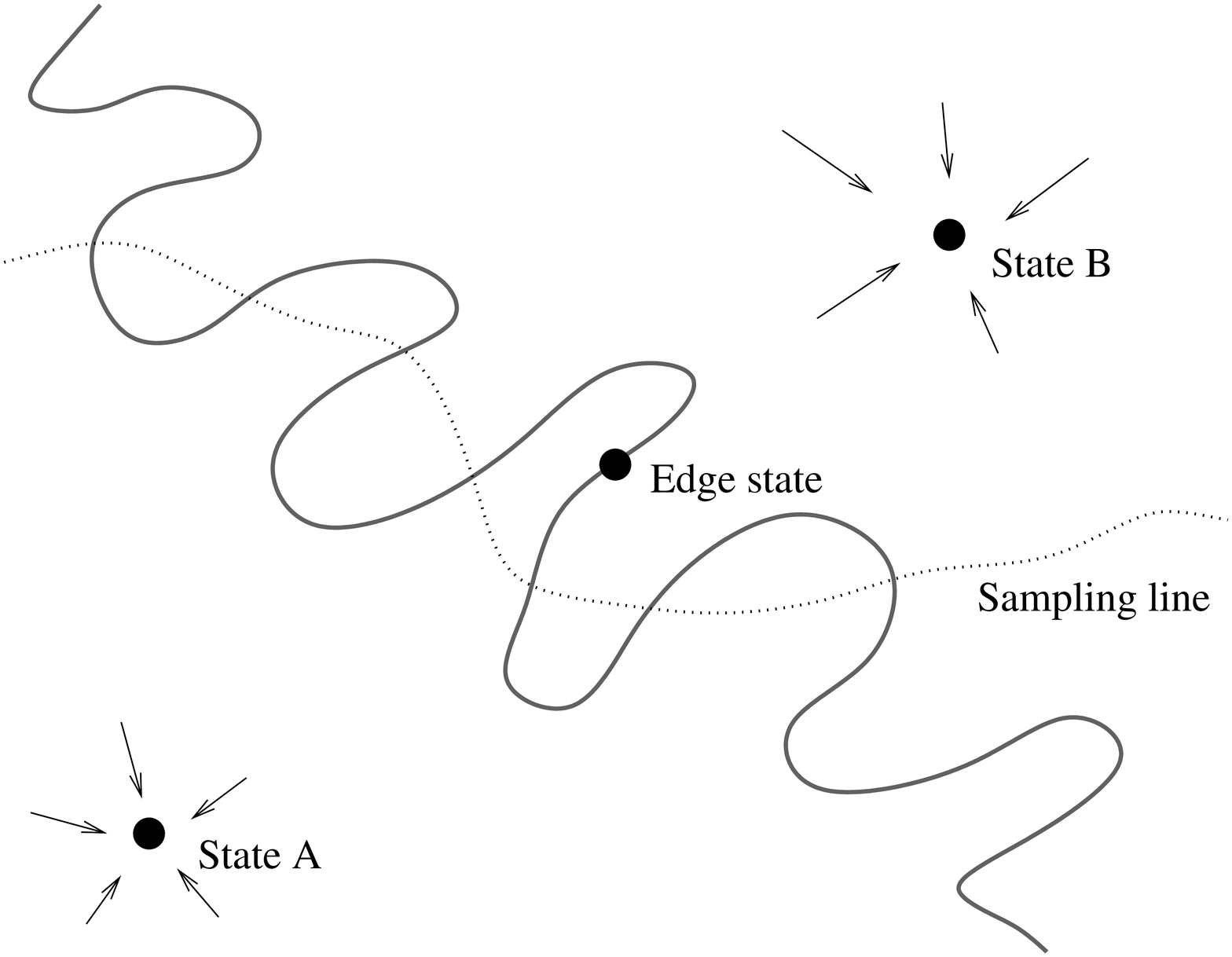}
\caption{Sketch of the phase space structure in the subcritical regime. The idea for this figure is taken from Ref. \cite{Skufca2006} and has been adapted to the present case.}
\label{Phase}
\end{center}
\end{figure}

A picture of the phase space structure in the subcritical case is illustrated in Fig. \ref{Phase}. Shown are the original equilibrium state A, the steady moist convection state B and the edge state. The figure indicates also the complex interface between both basins of attraction that we detected in several ways in the present study.

\subsection{Supercritical regime}
\begin{figure}
\begin{center}
\includegraphics[angle=0,scale=0.6,draft=false]{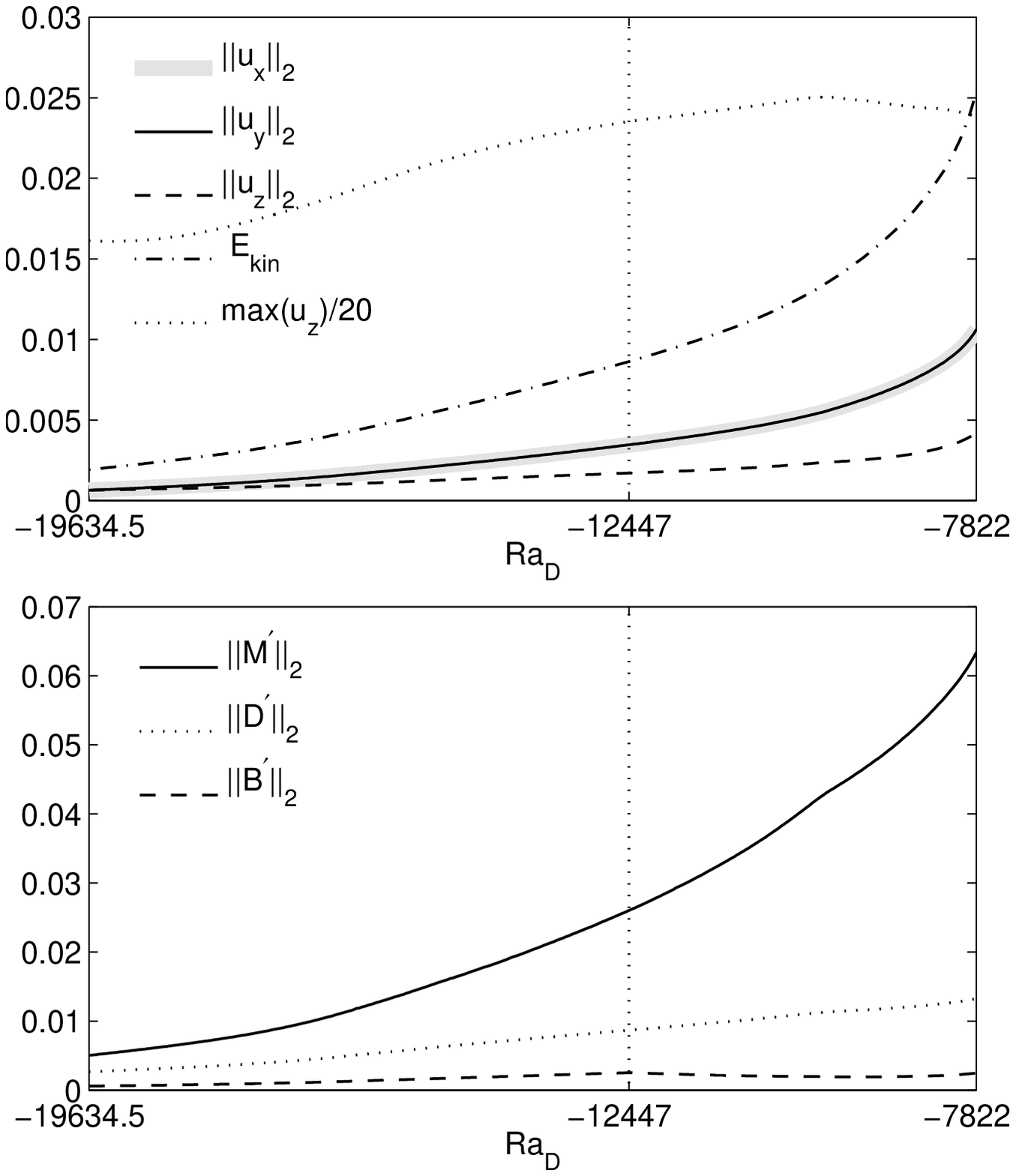}
\caption{Dependence of several fields on $Ra_D$ for convection in steady state B in the cell with $AR=4$. The dotted vertical line marks the KB case right between subcritical and supercritical regimes. We take the $L_2$-norm of the fields which is given by  $\parallel\cdot\parallel_2 =1/V \int_V |\cdot|^2 d {\bf x}$. For the whole range of $Ra_D$ convection is still in a stationary regime.}
\label{Dependence}
\end{center}
\end{figure}
\subsubsection{Structure of convection states in the supercritical regime}
\begin{figure}
\begin{center}
\includegraphics[angle=0,scale=0.6,draft=false]{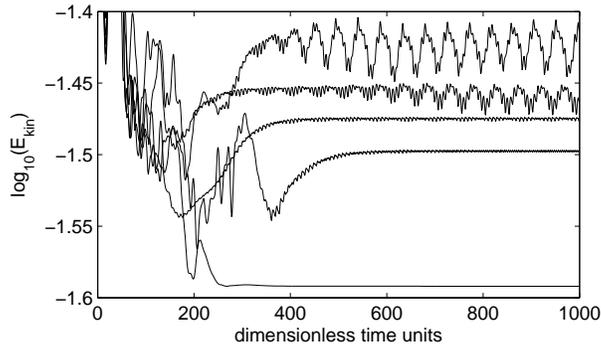}
\caption{Transition from stationary to time-dependent moist convection. The dry Rayleigh number $Ra_D$ is therefore increased from -7840 up to -7540 in increments of 60. Data are for $AR=4$, $SSD=0$, $CSA=4/3$, and $Ra_M=3.73\times 10^4$.} 
\label{TimeDep}
\end{center}
\end{figure}

In the supercritical regime, the diffusive equilibrium is saturated and linearly unstable.  A convective state can always be initiated. This was already shown in Fig. \ref{Transition} for aspect ratios 3 (right) and 4 (left). Typical convection states B in the sub- and supercritical regime are compared in Fig. \ref{Bsp4}. One can see that the shape of the moist plume changes qualitatively. While in the subcritical regime an isolated cloud exists, in the supercritical regime a closed cloud layer at the top is found. For decreased stratification the moist updraft becomes stronger while the overall shape remains the same. In Fig. \ref{Dependence} (top), the $L_2$ norms of the velocity components as a function of $Ra_D$ are shown. The two horizontal velocity magnitudes are the same as one would expect due to the symmetry. All three contributions to the kinetic energy grow steadily until state B becomes non-stationary in the supercritical regime. This is observed at about $Ra_D=-7820$ (see Fig. \ref{TimeDep}). Interestingly, the maximal vertical velocity inside the cloud varies much less for the range of $Ra_D$. The bottom panel of the same figure displays the $L_2$ norms of moist and dry buoyancy. The moist updraft increases in intensity as we reduce the stable stratification which is indicated by  $||M^{\prime}||_2$. The shrinking of the buoyancy fluctuation at the KB case is due to the fact, that the background equilibrium switches from fully unsaturated  to fully saturated case. Note also that the moist  Rayleigh number $Ra_M$ remains constant in this study.
\subsubsection{Recharge-discharge convection}
\begin{figure}
\begin{center}
\includegraphics[angle=0,scale=0.6,draft=false]{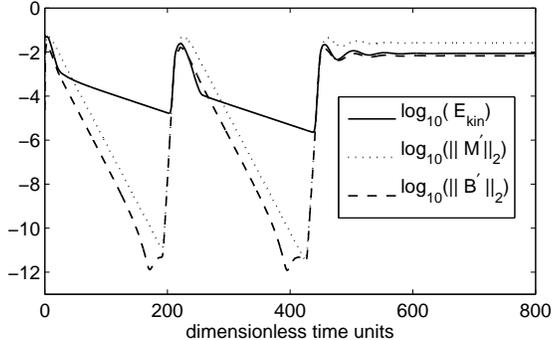}
\caption{Example for the recharge-discharge convection for $AR=4$. The initial equilibrium state is fully saturated and very close to the KB case. When time axis is rescaled by the diffusion time of the flow, $t_d=H^2/\nu$, the two recharging events will appear at $t/t_d=1$ and 2.}
\label{Recharge}
\end{center}
\end{figure}

For dry Rayleigh numbers in the supercritical regime right above the KB case, we observe in some instances moist convection in a highly intermittent recharge-discharge regime which is displayed  in Fig. \ref{Recharge} via the graphs of the $L_2$ norms of the velocity and the buoyancies versus time. After an initial convective burst of activity, the system slowly relaxes toward the quiescent equilibrium. However, after a while, convection is reinitiated and decays again. Interestingly, a similar behavior has been found in a channel flow with a strong transverse magnetic field \cite{Boeck2008} that enforces a switching between two-dimensional and three-dimensional flow states. The reason of why the conditionally unstable moist convection is not fully relaxing to the quiescent state can be explained as follows. In the supercritical regime, the equilibrium state is unstably stratified. Small perturbations can trigger new moist convection. We will get back to this discussion at the end of this section.

The number of such cycles can vary from one to several. An observed trend is that the closer the parameter set is to the KB parameter set the more cycles are possible. In the Galerkin model, it is found that the recharge-discharge regime also depends on the two Rayleigh numbers. For higher Rayleigh numbers, more cycles can be observed, which implies that viscosity plays an important role for this phenomenon. In particular,  the duration of the cycle is approximately equal to the diffusive time-scale $t_{d}=H^2/\nu$. This indicates that the quiescent relaxation is associated with the  diffusion of dry and stably stratified air across the entire layer. However, as the origin is a linearly unstable equilibrium, the quiescent relaxation cannot be sustained indefinitely. It is terminated by the abrupt onset of a cloudy moist plume. This behavior has been reproduced in direct numerical simulations (DNS) which will be reported elsewhere \cite{Pauluis2011}.

\begin{figure}
\begin{center}
\includegraphics[angle=0,scale=0.6,draft=false]{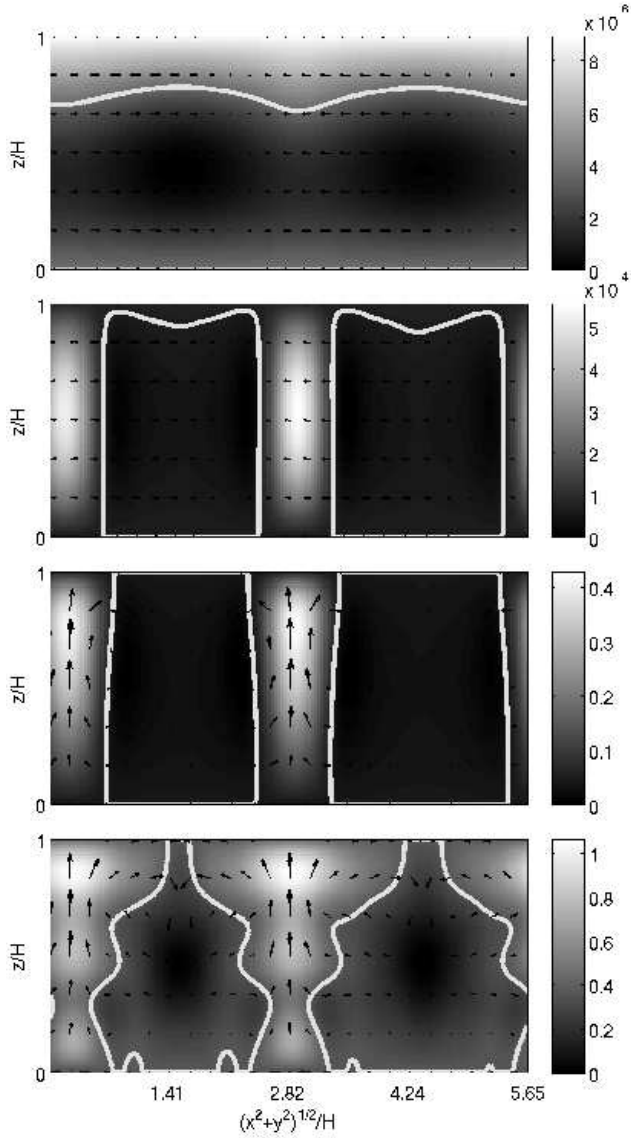}
\caption{Snapshots of the liquid water content $q_l$ and the projection of the velocity field into the plane for the same data as in Fig. \ref{Recharge}. They are taken in the plane with angle $\pi/4$ to the $x-z$ and $y-z$ plane. Bright line marks $q_l=0$. From top to bottom, first panel is at $t=183$. Velocity vectors are enhanced by a factor of 250. Second panel is at  $t=198$ and  velocity enhanced again by 250. Third panel is taken at $t=210$. The velocity vectors are now stretched by a factor of 6. Fourth panel taken at  $t=224$, velocity is enhanced by a factor of 4. Note that the colorbars of the panels differ by orders of magnitude.}
\label{Discharge}
\end{center}
\end{figure}

We illustrate this dynamics in detail for an example in Figs. \ref{Recharge} and \ref{Discharge}. Different phases of the evolution are displayed. The cycle starts at $t=27$ and is characterized by an exponential decay of all quantities (see Fig. \ref{Recharge}). At that time, the entire box is filled with unsaturated air and is stably stratified, except in the vicinity of the upper plate which is weakly saturated due to the boundary conditions. Since the system converges to the quiescent equilibrium state, moist air starts filling slowly the box from the upper plate. This process begins at  $t=171$ where the buoyancy $B^{\prime}$ is not decreasing any further because of the occurrence of moist buoyant air. The first panel of Fig. \ref{Discharge} taken at $t=183$
illustrates this stage slightly before the onset of convection. Fluid motions are very weak, and mostly horizontal. Saturated air begins to fill the layer slowly from above. At time $t= 191$, the buoyancy is increasing rapidly and the moist air in the box is rearranged to column-like clouds as shown in the second panel for $t=198$. The solid bright line is again the cloud boundary. The velocity remains small. But at $t= 203$ the buoyancy is strong enough to form moist upward motion in the cloud columns (see the third panel at $t=210$). Due to latent heat release this process is self-amplifying. The velocity increases very rapidly and the cloud columns form intense single updrafts up to $t= 224$ (fourth panel of Fig. \ref{Discharge}). If the updrafts are too strong, the convection can over-stabilize the layer which then must be destabilized by diffusion again. This means that a new recharge-discharge cycle is initiated. Alternatively, weaker updrafts may lead to a sustained convective regime in which the stabilization by convection balances diffusion on the same time-scale. In general, the larger the domain the easier to escape this cyclic recharge-discharge convection.

\begin{figure}
\begin{center}
\includegraphics[angle=0,scale=0.6,draft=false]{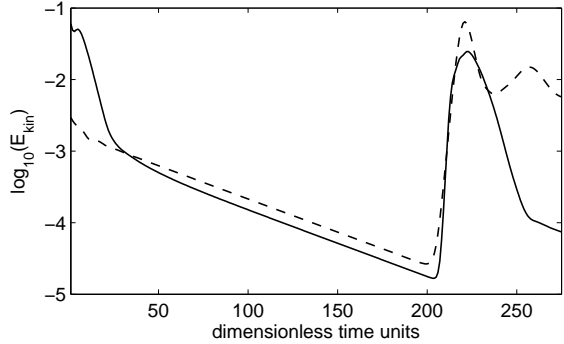}
\caption{Effect of mode reduction on the recharge-discharge convection regime. 
Solid line is for the usual resolution in our Galerkin approximation, dashed line 
for the reduced resolution.} 
\label{Reduced}
\end{center}
\end{figure}

In order to detect which Galerkin modes are mainly included in the recharge-discharge 
dynamics, we reduced the wavenumber space stepwise. It was possible to reproduce the cycles 
with wavenumbers $(|n_x| \le 1 , |n_y|\le 1 , | n_z| \le 2)$ (see Fig. \ref{Reduced}). We can thus 
conclude that this regime of convection is dominated by the large-scale degrees of freedom in the 
flow model.
\begin{figure}
\begin{center}
\includegraphics[angle=0,scale=0.28,draft=false]{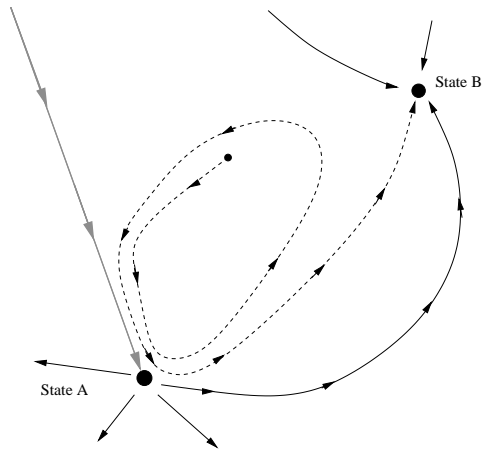}
\caption{Sketch of the phase space structure which is associated with
the recharge-discharge behavior. The gray arrow marks the case of a perturbation only 
in the $u_x$ and $u_y$ direction. The dashed line is a example for a two cycle recharge-discharge 
run.}
\label{Phase2}
\end{center}
\end{figure}

The conditions for the onset of a recharge-discharge cycle can be understood as follows (see also Fig. \ref{Discharge}). First, nearly the whole box must be filled with dry and stable air. Second, the velocity field is then almost horizontal, because an upward motion could create moist and unstable air and thus more upward motion. If we take such an initial condition for less stably stratified dry air (i.e. a lower $Ra_D$) we can produce at least one cycle. For parameter sets close to the KB case it is easier to satisfy these conditions. A small perturbation in the dry buoyancy field is enough to get an almost dry domain. That is why we can observe this phenomenon mostly close to the KB case. The closer we are to the KB case the more cycles are observable. In Fig. \ref{Phase2} we give a picture for the phase space structure for the recharge-discharge convection. It is clear from the equations of motion that perturbations in $u_x$ and $u_y$ only will always relax to state A. This path corresponds with a stable manifold and is indicated by the gray arrow in Fig. \ref{Phase2}.
\section{Summary and discussion}
The transition to turbulence in moist Rayleigh-B\'{e}nard convection starting from a conditionally unstable (CU) equilibrium has been studied in a  Fourier-Galerkin model. We classified the initial CU equilibrium of the layer in two different regimes, the subcritical unsaturated and the supercritical saturated. The transition behavior and the evolving dynamics are significantly altered by the degree of saturation which is controlled at the top of the layer by $Ra_D$ (when the parameters $SSD$, $CSA$, $Ra_M$ and $Pr$ are held fixed)  and the aspect ratio $AR$ of the layer.  Our study extends  
earlier results of Bretherton \cite{Bretherton1987,Bretherton1988} by demonstrating the existence of 
nonlinear three-dimensional convective regimes in a broad portion of the parameter space.

In contrast to classical dry RB case, convection is localized and does not fill the entire layer with rolls and thermal plumes. In the subcritical case, we showed that a narrow moist plume is surrounded by dry unsaturated air in which the fluid moves downward by diffusion (see Fig. \ref{Bsp2}).
This characteristic structure is well-known from observations and numerical simulations \cite{Bjerknes1938,Malkus1953,Siebesma1998} -- also from those of the KB regime 
\cite{Kuo1961,Bretherton1987,Bretherton1988}. In the supercritical regime, the top layer is however always saturated, a closed cloud layer is then fed by a narrow moist plume from below (see Fig. \ref{Bsp4}). The up-down symmetry as known from the dry RB case is consequently broken. Moist CU convection is additionally constrained by the aspect ratio. Time-dependent and eventually turbulent convection requires sufficiently extended cells. 

In the subcritical case, when the atmosphere remains subsaturated at the top ($\overline{M}(z) < \overline{D}(z) -N_s^2 z$), two stable attractors or fixed points coexist. The first one corresponds to a diffusive atmosphere with no motion, while the second one exhibits an overturning circulation in the form of  a steady rising moist plume (or cloud) balanced by subsidence in the unsaturated environment. The existence  of these multiple attractors has been confirmed through several statistical analyses. With decreasing $Ra_D$, it becomes less likely that a convective equilibrium  can be obtained. For a sufficiently negative dry Rayleigh number, that is however still well above the threshold for  which the layer is absolutely stable ($CAPE<0$), the probability to switch to moist convection becomes zero. In this range of $Ra_D$, we traced a periodic orbit at the edge of the basin of attraction of the convective state (see Fig. \ref{Phase}). Some of the transition properties in this regime are similar to those of wall-bounded shear flows. We mention the sensitive dependence on the amplitude of the finite perturbation and the complex shaped boundary between laminar and turbulent states. Furthermore the convective state is localized in space, very similar to turbulent spots in a shear flow \cite{Schumacher2001}. But there are also differences. In the present model, we did not observe transient convection states which suggests the existence of an attractor rather than a chaotic saddle (see Fig. \ref{Phase}). An exception is the edge state.

If the layer is held saturated at the top ($\overline{M}(z) > \overline{D}(z) -N_s^2 z$), the initial equilibrium is linearly  unstable with respect to small amplitude perturbations. The system switches into a moist convection mode which in dependence on the chosen aspect ratio and the degree of saturation at the top can be a recharge-discharge mode or a stationary moist plume mode. A further increase in the dry Rayleigh number $Ra_D$, which corresponds to a decreasing stratification in the unsaturated environment, causes the moist plume structure to become time-dependent and eventually turbulent.

A few more words are in order now. The present analysis is focused on  moderate Rayleigh number and small aspect ratios of 3 to 8. These choices, made necessary by the Galerkin truncation, will have their influence on the initialization of cloud patterns. Moderate Rayleigh numbers and thus larger viscosity and diffusivity ease the onset of sustained convective motion
outside the cloud since the stable dry air can descent faster there by diffusion. With increasing Rayleigh number this process becomes increasingly inefficient. In other words, the upward transport in ascending plumes must be balanced by slower subsidence occupying larger and larger portions of the domain around the clouds. Thus, bigger aspect ratios are likely to be necessary to obtain a self-sustained convective regime at high Rayleigh number. This is confirmed by recent high-resolution DNS of this model at higher Rayleigh numbers \cite{Pauluis2011}. Note also, that with a view to moist convection in the atmosphere, radiative cooling or non-equilibrium effects such as precipitation step in and can become important physical processes to establish moist convection in the conditionally unstable regime. They provide the additional dissipation mechanisms that amplify the downward fluid transport outside the moist plume when diffusive transport becomes increasingly inefficient with growing Rayleigh (or Reynolds) numbers and will ease the constraint of larger aspect ratios.

The results presented here can be expanded in several directions. First, our study only considered the case where the lower boundary is exactly at the saturation boundary, corresponding to  $SSD = 0$, which can be interpreted as convection over an ocean surface. Further exploration of state space to include cases with non-zero $SSD$ could shed some light on the different behavior between convection over land and over the oceans.  Second, changing boundary conditions from free-slip to constant flux conditions  can have an influence on the dynamics at moderate Rayleigh number as known from dry convection \cite{Johnston2009}. Third, the addition of a radiative cooling, which occurs in the atmosphere due to the emission of infrared radiation by greenhouse gases, will significantly enhance the number of dynamical regimes as has been recently discussed by large eddy simulations with a parametrization of small-scale turbulence in Ref. \cite{Feingold2010}.
\vspace{0.5cm}
\acknowledgments
The work of TW and JS is supported by the Deutsche Forschungsgemeinschaft (DFG) under grant no. SCHU1410/8-1 and the DFG Heisenberg Program under grant SCHU1410/5-1. OP is supported by the US National Science Foundation under grant ATM-0545047. We would also like 
to thank the J\"ulich Supercomputing Centre (J\"ulich, Germany) for computing time on the Blue Gene/P system JUGENE under grant HIL02 at which supercomputations were conducted that complemented the present work.

\end{document}